\documentclass[onecolumn,aps,prd,preprintnumbers,superscriptaddress,nofootinbib,amsmath,amssymb,floats,floatfix,showkeys,notitlepage,showpacs]{revtex4-1}

\usepackage{orcidlink}
\usepackage{comment}
\usepackage{lipsum}
\usepackage{graphicx}
\usepackage{subfigure}
\usepackage{palatino}
\usepackage{float}
\usepackage{sans}
\usepackage{hyperref}
\hypersetup{colorlinks=true,linkcolor=blue,urlcolor=blue,citecolor=blue}
\usepackage[toc,page]{appendix}
\usepackage[normalem]{ulem}
\usepackage{adjustbox}
\usepackage{latexsym}
\usepackage{amsmath}
\usepackage{breqn}
\usepackage{amssymb}
\usepackage{amsfonts}
\usepackage{dcolumn}
\usepackage{bm}
\usepackage{tikz}
\usetikzlibrary{shapes.geometric, arrows, positioning, arrows.meta}
\usepackage{pgfplots}
\pgfplotsset{compat=1.18}
\usepackage{bigints}
\usepackage{array,tabularx,multirow,booktabs}
\usepackage[tracking=true]{microtype}
\SetTracking{}{500}
\SetTracking{encoding={*}, shape=sc}{40}
\UseRawInputEncoding 
\allowdisplaybreaks

\begin{document}
	\newcommand \nn{\nonumber}
	\newcommand \fc{\frac}
	\newcommand \lt{\left}
	\newcommand \rt{\right}
	\newcommand \pd{\partial}
	\newcommand \e{\text{e}}
	\newcommand \hmn{h_{\mu\nu}}
	\newcommand{\PR}[1]{\ensuremath{\left[#1\right]}} 
	\newcommand{\PC}[1]{\ensuremath{\left(#1\right)}} 
	\newcommand{\PX}[1]{\ensuremath{\left\lbrace#1\right\rbrace}} 
	\newcommand{\BR}[1]{\ensuremath{\left\langle#1\right\vert}} 
	\newcommand{\KT}[1]{\ensuremath{\left\vert#1\right\rangle}} 
	\newcommand{\MD}[1]{\ensuremath{\left\vert#1\right\vert}} 

	

%
\title{\textcolor{blue}{QNMs of charged black holes in AdS spacetime: a geometrical optics perspective }}
\author{
Shubham Kala \orcidlink{0000-0003-2379-0204}}
\email{shubhamkala871@gmail.com}
\affiliation{The Institute of Mathematical Sciences$,$ C$.$I$.$T$.$ Campus$,$ Taramani$,$ Chennai\textminus600113$,$ India}

\author{Anik Rudra \orcidlink{0000-0002-6710-9778}}
\email{anikrudra23@gmail.com}
\affiliation{School of Physics and Mandelstam Institute for Theoretical Physics$,$ University of the Witwatersrand$,$ Wits$,$ 2050$,$ South Africa}

\author{Hemwati Nandan \orcidlink{0000-0002-1183-4727}}
\email{hnandan@associates.iucaa.in}
\affiliation{Department of Physics$,$ Hemvati Nandan Bahugunga Garhwal Central University$,$ Srinagar\textminus 246174$,$ Uttarakhand$,$ India\newline
Center for Space Research North\textminus West University$,$ Potchefstroom\textminus 2520$,$ South Africa}

\begin{abstract}
We investigate the quasinormal modes of the Reissner$-$Nordstr\"om anti$-$de Sitter black hole using the Penrose limit, motivated by the geometrical optics approximation. This approach offers a novel framework for approximating quasinormal modes with large real frequencies by associating a plane wave to spacetime regions near null geodesics, providing a geometric interpretation of the geometrical optics approximation. Applying this limit to bound null orbits around black holes allows us to explore the black hole response to perturbations. We analyze the effects of black hole charge and negative cosmological constant on the quasinormal spectrum, finding that increasing charge enhances both the real and imaginary parts of the frequencies, while a decreasing cosmological constant leads to higher real frequencies and longer lived perturbations, with the spectrum stabilizing at larger values of the cosmological constant.
\end{abstract}	



\maketitle


\section{Introduction}\label{intro}
The study of quasinormal modes (QNMs) around black holes (BHs) has become an important area of research in theoretical physics, providing deep insights into the dynamics of spacetime perturbations \cite{Chandrasekhar:1975zza}. QNMs represent the characteristic oscillations of a perturbed BH, where spacetime vibrates at complex frequencies, embodying a combination of oscillatory and damped behavior \cite{Kokkotas:1999bd,Berti:2009kk}. These modes are intrinsically connected to the stability and response of BHs under perturbations, making them a cornerstone for understanding gravitational waves and their interaction with the BH's geometry \cite{Sasaki:2003xr,Baker:2005vv,Ferrari:2007dd}. The study of QNMs has evolved significantly, transitioning from traditional analytical techniques to modern computational and numerical methods. Traditionally, QNMs have been explored using analytic approaches such as the WKB approximation (Wentzel-Kramers-Brillouin), which provides approximate solutions for perturbations in BH spacetimes \cite{Iyer:1986np}. These methods, while insightful, were often limited to specific conditions such as low multipole moments or small perturbations, and faced challenges to achieve high precision for complex spacetimes \cite{Daghigh:2011ty}. Recent advances have introduced more robust numerical techniques, including the use of finite-difference methods, spectral decomposition, and complex frequency domain analysis, which enable the precise computation of QNMs for a broader range of BH configurations, including those with higher spins, charges, or exotic features \cite{cho2012new,Matyjasek:2017psv,Gonzalez:2017gwa,Bizon:2020qnd,lopez2020numerical,Roussille:2023sdr,heidari2023investigation,roussille2024numerical,PanossoMacedo:2024nkw}. A new method has been developed to study the dynamics of BHs by focusing on the photon ring \cite{Fransen:2023eqj}, a region defined by nearly bound null geodesics that encode critical information about the geometry and properties of the BH \cite{darwin1959gravity}. This approach leverages the Penrose limit, a space-time simplification technique that transforms the complex near-photon ring dynamics into a plane wave metric \cite{hadar2022holography,blau2011plane}. By isolating the physics of the photon ring while preserving essential features like geodesic deviation and tidal forces, this method provides a robust framework for investigating the intricate behavior of BH spacetimes \cite{cardoso2009geodesic,eisenhart1928dynamical}. Furthermore, it establishes a foundation for rigorous investigation of the intricate connections between the dynamics of the photon ring and phenomena such as QNMs, thus providing a deeper understanding of gravitational physics and its observational consequences \cite{hod2009black,dolan2010quasinormal,yang2012quasinormal}.\\
QNMs in anti-de Sitter (AdS) spacetime were first computed for a conformally invariant scalar field, whose asymptotic behavior is similar to that in flat spacetime \cite{Chan:1996yk}. Later, motivated by the AdS-CFT (conformal field theory) correspondence, a systematic computation of QNMs was carried out for scalar perturbations of Schwarzschild-AdS (S-AdS) spacetimes \cite{Horowitz:1999jd}. This work was subsequently extended to include gravitational and electromagnetic perturbations of S-AdS BHs \cite{Cardoso:2001bb}. Reconsideration of the appropriate boundary conditions for the QNM problem led to further insights into gravitational perturbations of both Schwarzschild-dS and AdS spacetimes \cite{Moss:2001ga}. Finally, the study of scalar perturbations was expanded to encompass Reissner-Nordstr\"om-AdS (RN-AdS) BHs \cite{Wang:2000gsa}. Further, studies of dS and AdS BH spacetimes have incorporated time evolution methods alongside traditional frequency-domain analyses. Time evolution techniques have been used to investigate the late-time behavior of perturbations in Schwarzschild-dS backgrounds from a cosmological perspective \cite{Brady:1996za,Brady:1999wd}. In the context of the AdS/CFT correspondence, similar time evolution studies have also been conducted for RN-AdS spacetimes \cite{Wang:2000dt}. Numerical approaches to computing QNMs in AdS backgrounds often encounter difficulties in the Schwarzschild-AdS case when the BH horizon is significantly smaller than the AdS radius \cite{Horowitz:1999jd}. As a result, particular emphasis has been placed on studying the small BH regime, utilizing both frequency-domain and time-domain methods to improve accuracy and stability \cite{Konoplya:2002zu,Zhu:2001vi}. Berti et al. presented a comprehensive analysis for scalar, electromagnetic, and gravitational perturbations of RN-AdS BH in the frequency domain \cite{berti2003quasinormal}. Their study verified, refined, and in some cases challenged previous findings while extending existing results. Additionally, electromagnetic and gravitational QNMs for RN-AdS BHs were computed for the first time in this study, drawing inspiration from prior investigations of RN-dS spacetimes in a cosmological context. With the increasing significance of this field, numerous studies have investigated QNMs in AdS geometries within both GR and alternative theories of gravity, yielding important insights and advancements, some of which are referenced herein \cite{Yan:2020nvk,Yan:2020hga,Fontana:2020syy,Mourier:2020mwa,Chen:2020evr,Jusufi:2020odz,Konoplya:2020kqb,Rincon:2020pne,Aragon:2020xtm,Okyay:2021nnh,Capano:2021etf,Pierini:2021jxd,Srivastava:2021imr,Pantig:2022gih,Rayimbaev:2022mrk,Pierini:2022eim,Atamurotov:2022nim,Gogoi:2022wyv,Zhao:2022gxl,Konoplya:2022iyn,Yang:2022xxh,Konoplya:2022tvv,Konoplya:2022zav,Baibhav:2023clw,Lambiase:2023hng,Gogoi:2023kjt,Konoplya:2023aph,Gong:2023ghh,Gogoi:2023fow,Ghosh:2023etd,Filho:2023ycx,Baruah:2023rhd,Chen:2023cjd,Vishvakarma:2023csw,Ma:2024qcv,Pedrotti:2024znu,Stashko:2024wuq,Konoplya:2024hfg,Konoplya:2024vuj,Fatima:2024gji,Konoplya:2024ptj,Karmakar:2024xwr,AraujoFilho:2024lsi,Jia:2024pdk,Ge:2025xuy}.\\
Our aim in this paper is to study the QNMs around a RN-AdS BH using a recently developed method that focuses on the photon ring. This approach utilizes the Penrose limit to simplify the near-photon ring dynamics, transforming them into a plane wave metric while preserving key features like geodesic deviation and tidal forces. In Section \ref{sec2}, we introduce the RN AdS BH spacetime and outline its key properties. Section \ref{sec3} provides a brief review of plane waves and the Penrose limit, which allows us to approximate the BH geometry near null geodesics. Building on this, Section \ref{sec4} explores the application of the Penrose limit to bound null orbits, offering a geometric perspective on perturbations around the RN AdS BH. In Section \ref{sec5}, we discuss Lyapunov exponents and orbital frequencies, which are crucial for understanding the stability of geodesic motion and its relation to QNMs. Section \ref{sec6} presents our main results on the QNM spectrum, analyzing how BH charge and the cosmological constant influence the oscillatory and damping behavior of perturbations. Finally, in Section \ref{sec7}, we briefly summarize our findings.
\section{RN AdS BH Spacetime} \label{sec2}
We consider the line element of Reissner-Nordstr\"{o}m AdS BH in spherical coordinates as \cite{Louko:1996dw},
\begin{equation} \label{metric}
    ds^{2} = f(r) dr^{2} -\frac{dr^2}{f(r)} - r^{2} (d\theta^{2}+\sin^{2}\theta d\phi^{2}),
\end{equation}
where, 
\begin{equation}
    f(r) = \frac{\Delta(r)}{r^{2} \ell^{2}} \And \Delta(r) = r^{4} + \ell^{2}(r^{2}-2Mr+Q^{2}).
\end{equation}
Here, $M$ is the mass of BH and $Q$ represents the electric charge of BH and $\ell$ is a cosmological constant parameter ($\Lambda=-3/\ell^{2}$).
The massive scalar wave equation corresponding to metric \eqref{metric} can be written as follows \cite{Zamorano:1982zz},
\begin{equation}
    \frac{1}{\sqrt{-g}}\partial_{\mu} \left( \sqrt{-g}g^{\mu\nu}\partial_{\nu}\psi \right), \hspace{0.5cm} g = det(g_{\mu\nu}).
\end{equation}
The massive scalar wave equation can be reduced to its radial components as,
\begin{equation}
    \frac{d^2\psi_s}{dr^2_*} + (\omega^2 - V_s(r_{*}))\psi_s = 0.
\end{equation}
Here, $r_{*}$ is the tortoise coordinate associated with $dr_* = \frac{dr}{f(r)}$ and given as,
\begin{equation}
    dr_* = \frac{dr}{\left( 1-\frac{2M}{r}+\frac{Q^2}{r^2} +\frac{r^{2}}{\ell^{2}} \right)}.
\end{equation}
Therefore, the scalar effective potential is given as follows\footnote{Please note the distinction between the symbols $l$ and $\ell$. The symbol $l$ refers to the angular quantum number associated with the scalar wave function. On the other hand, throughout the paper, we use $\ell$ as a parameter related to the cosmological constant, as introduced earlier.},
\begin{equation}
    V_{scalar}(r_{*}) =  \left( 1-\frac{2M}{r}+\frac{Q^2}{r^2} +\frac{r^{2}}{\ell^{2}} \right) \left( \frac{l(l+1)}{r^2} +\frac{2M}{r^{3}} - \frac{2Q^2}{r^4} + \frac{2r}{\ell^{2}} \right).
\end{equation}
Since this work employs a different method to compute the QNMs, a detailed exploration of the scalar potential is not necessary. Therefore, we only provide its expression without further expansion.
\section{Brief Review of Plane Waves and Penrose Limit} \label{sec3}
The geometrical optics approximation provides a powerful method to analyze wave propagation in curved spacetimes. In this approach, the wave equation is approximated by considering high-frequency limits where wavefronts follow null geodesics \cite{blau2006fermi}. Specifically, when applying the geometrical optics approximation to a wave equation on a spacetime \((M, g)\), the evolution of the leading amplitude along a null geodesic \(\gamma\) governs the wave behavior \cite{hod2009black}. Remarkably, this approximation aligns exactly with the full-wave equation formulated on the Penrose limit spacetime \((M_{\gamma}, g_{\gamma})\). The Penrose limit effectively zooms in on the geometry near a given null geodesic, simplifying the background spacetime into a plane wave metric \cite{Kapec:2022dvc}. This correspondence suggests that, within the geometrical optics regime, studying wave equations on the original curved spacetime can be equivalently understood through their exact counterparts in the Penrose limit spacetime, thus providing deeper insights into wave propagation under extreme gravitational conditions.
\begin{figure}[H]
\begin{center}
\begin{tikzpicture}[
    node distance=2.5cm and 3.5cm,
    every node/.style={align=center, font=\normalsize},
    every path/.style={draw, thick, -{Latex}, shorten >=2pt} 
]

    \definecolor{myblue}{RGB}{70,130,180}   
    \definecolor{mygreen}{RGB}{34,139,34}   
    \definecolor{myred}{RGB}{220,20,60}     

    \node[rectangle, draw=myblue, thick, rounded corners, fill=blue!10, text width=5cm, minimum height=1cm] 
        (A) {General spacetime \\ $(M, g)$};
    
    \node[rectangle, draw=mygreen, thick, rounded corners, fill=green!10, text width=5cm, minimum height=1cm, right=of A] 
        (B) {Plane wave spacetime \\ $(M_{\gamma}, g_{\gamma})$};
    
    \node[rectangle, draw=myred, thick, rounded corners, fill=red!10, text width=5cm, minimum height=1cm, below=of A] 
        (C) {Wave equation on \\ $(M, g)$};
    
    \node[rectangle, draw=myred, thick, rounded corners, fill=red!10, text width=5cm, minimum height=1cm, below=of B] 
        (D) {Wave equation on \\ $(M_{\gamma}, g_{\gamma})$};

    \path (A) edge[->] node[above] {\textbf{Penrose limit}$_{\gamma}$} (B);
    \path (A) edge[->] (C);
    \path (B) edge[->] (D);
    \path (C) edge[->] node[above] {\textbf{Geometric optics}$_{\gamma}$} (D);
\end{tikzpicture}
\end{center}    
\caption{The schematic diagram illustrates the geometrical optics approximation. Here the leading-order amplitude evolution of the wave equation along a null geodesic $\gamma$ in a spacetime $M$ with metric $g$ corresponds exactly to the wave equation in the Penrose limit spacetime $M_{\gamma}$ with metric $g_{\gamma}$ \cite{Fransen:2023eqj}.}

\end{figure}
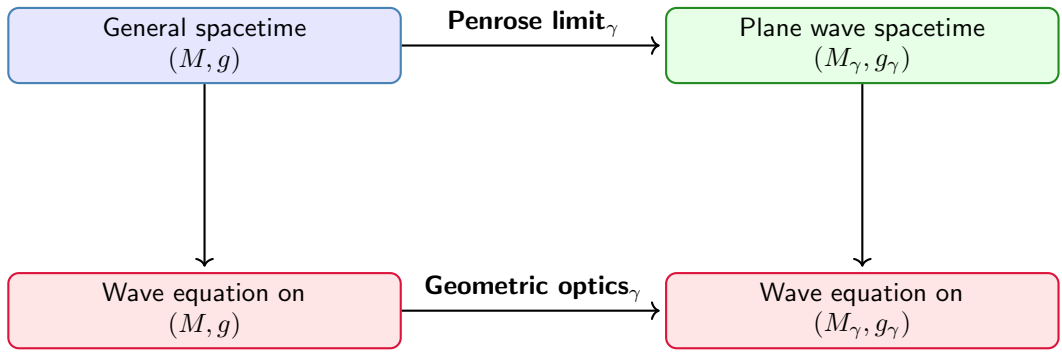

A null Fermi normal coordinate system adapted to the geodesic provides a natural framework for analyzing QNMs within the geometrical optics approximation. In this context, the dominant contribution to the spacetime structure in these coordinates corresponds to a plane wave spacetime. In other words, the physics governing the "near-photon ring" region can be effectively described by the Penrose limit of the original spacetime. It captures the leading-order geometry around a chosen null geodesic and formulated as \cite{penrose1976any},
\begin{equation}
     ds^{r} = 2dudv + A_{ij} (u) x^{i}x^{j} du^{2} + dx_{i} +dx^{i},
\end{equation}
Here, the Latin indices $i, j$ run over the transverse directions of the null Fermi-Walker coordinate system. The quantity $A_{ij}(u)$ is defined as  $A_{ij}(u) = R_{\mu i \nu j} u^\mu u^\nu$, where $ R_{\mu i \nu j}$ is the Riemann tensor and $u^\mu$ is the tangent vector to the null geodesic. This tidal tensor, evaluated along the null geodesic, generally depends on the affine parameter $u$ and governs geodesic deviation properties, including the Lyapunov exponent. These properties, in turn, influence the behavior of the associated QNMs in the geometrical optics approximation.
\section{Plane Waves and Penrose limit in RN A\MakeLowercase{d}S BH Spacetime} \label{sec4}
The plane wave solution in Brinkmann coordinates exhibits near uniqueness, constrained by the permissible coordinate transformations that preserve the imposed gauge conditions and the specific structure of the metric. In this construction, plane waves are derived by applying the Penrose limit to a chosen spacetime, focusing on null geodesics that constitute the photon ring. The Penrose limit captures the regime near the photon ring, and the emergent symmetries of the photon ring dynamics become manifest through the resulting plane wave geometry.\\
The Penrose limit plane wave in Brinkaman coordinates is given by \cite{penrose1976any,blau2011plane},
\begin{equation}
    ds^{r} = 2dudv + A_{ij} x^{i}x^{j} du^{2} + dx_{1}^2 +dx_{2}^2,
\end{equation}
where,
\begin{equation}
    A_{ij} = \left( R_{\mu\nu\alpha\beta} u^{\mu} e^{(i)\nu} u^{\alpha} e^{(j)\beta} \right)_{\gamma}.
\end{equation}
The plane wave metric depends almost uniquely on the symmetric matrix $A(u)$. This uniqueness arises because it is challenging to find a coordinate transformation that preserves the plane wave metric's form while satisfying the chosen gauge conditions. Consequently, the matrix $A_{ij}$ in Brinkmann coordinates assumes a distinct significance. Next, we need to construct the parallel frames for the null geodesics such that,
Now, we have to construct the parallel frames for the null geodesics such that \cite{Kubiznak:2008zs},
\begin{equation}
    u_{\mu} = \partial_{\mu} s, u_{\mu} u^{\mu} = n_{\mu} n^{\mu} =0, e_{\mu}^{(i)} e_{\mu}^{(j)} = \delta^{ij},
\end{equation}
where,
\begin{align}
e^{(1)\mu} &= \frac{1}{C} \left( u^{\alpha} h_{\alpha}^{\mu} -  u (u^{\alpha}S_{\alpha}) u^{\mu} \right), \\
e^{(1)\mu} &= \frac{1}{K} (u^{\alpha} Y_{\alpha}^{\mu}), \\
   n^{\mu} &= \frac{1}{C} e^(1)\alpha h_{\alpha}^{\mu} + \frac{1}{2C^{\mu}} \left( C_{\beta}^{\gamma} C_{\gamma\delta} u^{beta}u^{\delta} +u^{2} (\zeta_{\alpha} u^{\alpha})^{2} C^{2} \right)u^{\mu}.
\end{align}
Here, we have defined the conformal killing-yano tensor $h_{\mu\nu}$ as the Houdge dual of the killing-yano tensor $(Y)$ as \cite{Frolov:2008jr},
\begin{equation}
    h = Y = h_{\mu\nu} dx^{\mu}x^{\nu}.
\end{equation}
Hence, the simplified expression of $A_{ij}$ can be expressed as follow \cite{Giataganas:2024hil},
\begin{align} \label{Aij}
    A_{11} &= \frac{4}{r_{0}^2} - \frac{8 \Delta(r_{0}) [r_{0} \Delta^{''}(r_{0})-\Delta^{'}(r_{0})]}{r_{0}^{3}  \Delta^{'}(r_{0})^{2} } \\
    A_{22} &= \frac{-2}{r_{0}^2} +  \frac{8 \Delta(r_{0}) [\Delta^{'}(r_{0})-2r_{0}]} {r_{0}^{3}  \Delta^{'}(r_{0})^{2} } \\  
    \And
    A_{21} &= A_{12} = 0 .
\end{align}
For RN AdS BH,
\begin{equation}
    \Delta(r) = r^{4} + \ell^{2}(r^{2}-2Mr+Q^{2}).
\end{equation}
The BH horizon radii is therefore given by \cite{Cruz:2011yr}\footnote{For a detailed analysis of the horizon structure, please refer to the paper.},
\begin{equation}
    r_{h\pm} =  \mathcal{R}_{+} \pm \ell \left[ \frac{M}{2 \mathcal{R}_{+}} - \frac{ \mathcal{R}_{+}^{2}}{\ell^{2}} -\frac{1}{2}. \right].
\end{equation}
where,
\begin{equation}
    \mathcal{R}_{+} = \frac{\ell}{\sqrt{6}} \left[ \sqrt{1+\frac{12Q^{2}}{\ell^{2}}} \cosh \left[ \frac{1}{3} \cosh^{-1} \left( \frac{4+54M^2/\ell^2 -3(1+12Q^2/\ell^2)}{1+12Q^2/\ell^2}\right)\right] -1\right]^{1/2}.
\end{equation}
The radius of the photon sphere is given by \cite{Ladino:2024ned},
\begin{equation}
    r_{p\pm} = \frac{3M \pm \sqrt{9M^{2} -8Q^{2}}}{2}.
\end{equation}
From the expression of the photon sphere, it is evident that it is independent of the cosmological constant. Moreover, the photon sphere decreases with an increasing charge parameter. For $Q=M$ the photon sphere still exists but reaches its minimum value. Thus, even in the extremal case, the photon sphere does not vanish completely but reduces to a critical radius.
\section{Lyapunov Exponents and Orbital Frequencies} \label{sec5}
\begin{figure}[H]
	\begin{center}
        {\includegraphics[width=0.48\textwidth]{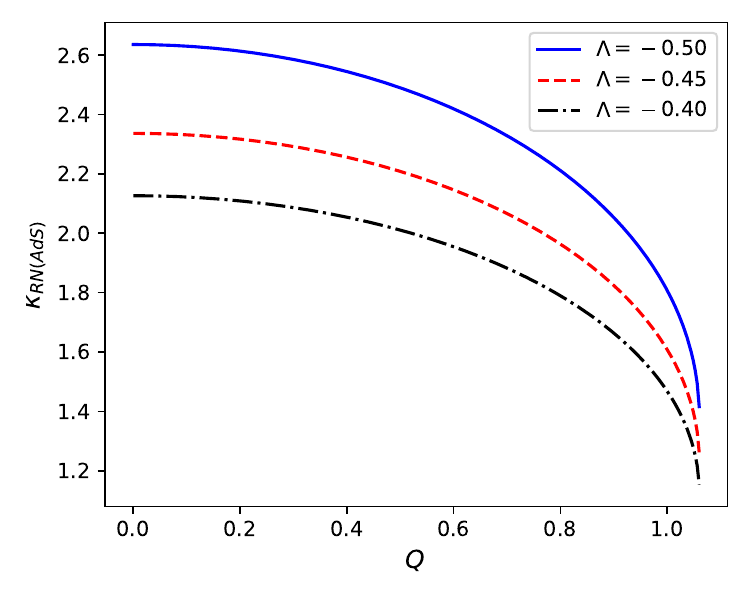}} 
	\end{center}
	\caption{The variation of surface gravity with charge parameter for different values of cosmological constant parameter with $M=1$.} \label{kappa}
\end{figure}

\begin{figure}[H]
	\begin{center}
        {\includegraphics[width=0.48\textwidth]{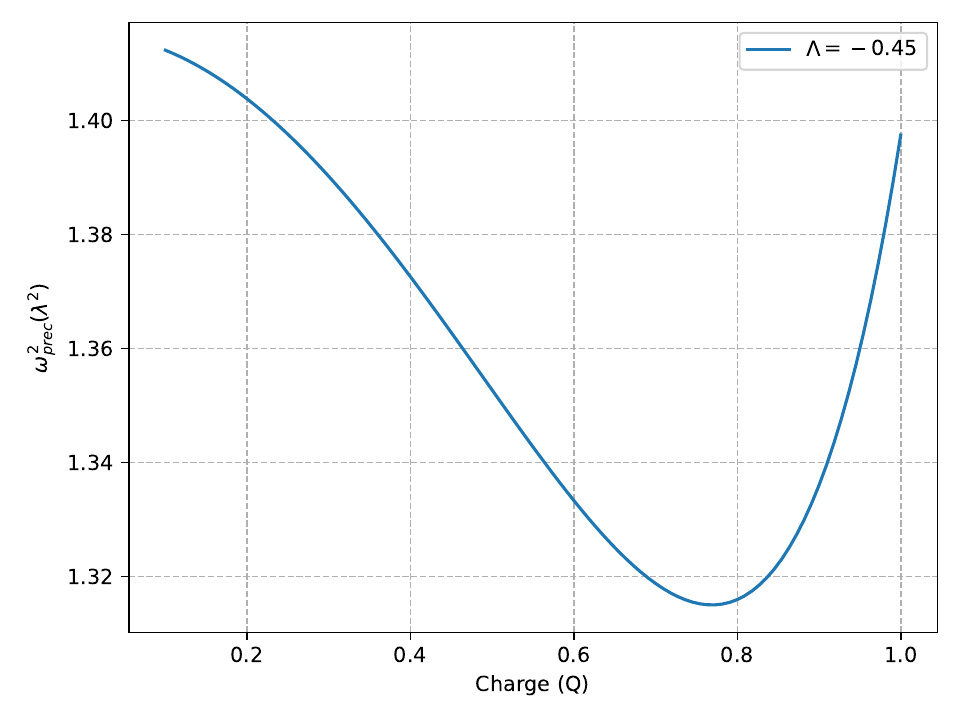}} 
        {\includegraphics[width=0.48\textwidth]{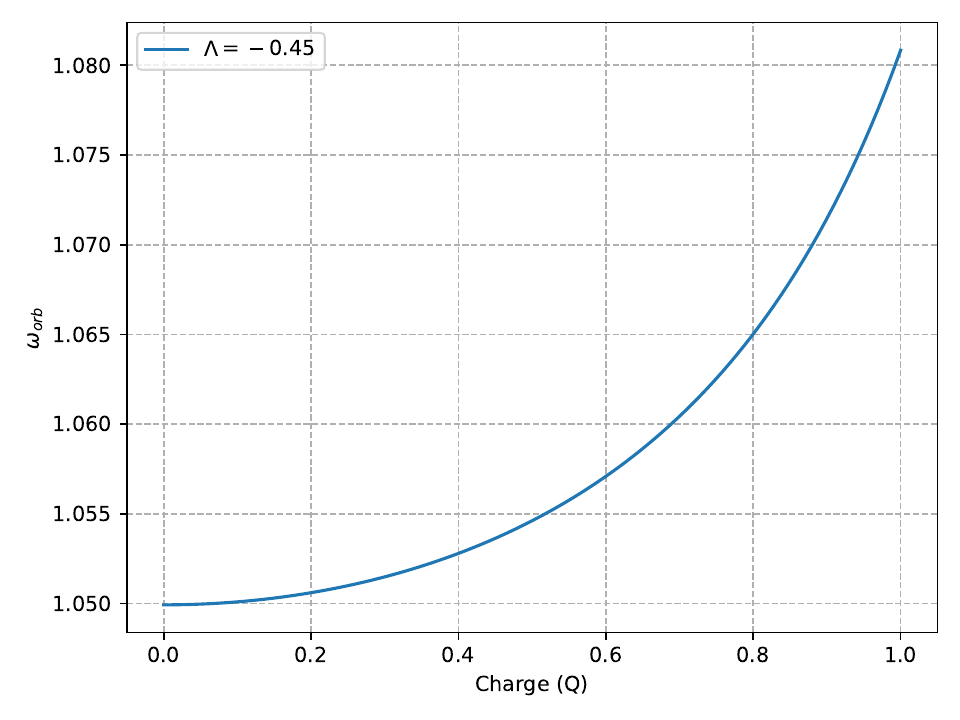}} 
	\end{center}
	\caption{The variation of precession (left) and orbital frequency (right) with electric charge parameter of BH with $M=1$ and $\Lambda=-0.45$.} \label{LvsQ}
\end{figure}

Now, by using the obtained expression in Eq. \eqref{Aij}, the metrics components for Penrose limit plane wave can be given as,
\begin{align}
    A_{11} &= \frac{4}{r_{p}^{2}} -\frac{8(r_{p}^{4} +\ell^{2}r_{p}^{2}-2M\ell^{2}r_{p}+Q^{2}\ell^{2})[r_{p}(12r_{p}^{2}+2\ell^{2})-4r_{p}^{3}+2\ell^{2}r_{p}-2M\ell^{2}]}{r_{p}^{3}(4r_{p}^{3}+2\ell^{2}r_{p}-2M\ell^{2})^{2}},\\
    A_{22} &= -\frac{2}{r_{p}^{2}} + \frac{8(r_{p}^{4} +\ell^{2}r_{p}^{2}-2M\ell^{2}r_{p}+Q^{2}\ell^{2}) [4r_{p}^{3}+2\ell^{2}r_{p}-2M\ell^{2}-2r_{p}]}{r_{p}^{3}(4r_{p}^{3}+2\ell^{2}r_{p}-2M\ell^{2})^{2}},
\end{align}
with $A_{12} = A_{21}=0$ and $A_{11} \neq -A_{22}$ always, unless $Q=0$.\\
The Lyapunov exponent of unstable null geodesics around near-extremal de-Sitter BHs and the imaginary part of the corresponding QNMs are observed to be related to the surface gravity of the BH \cite{Cardoso:2008bp}. This observation suggests a potential connection between the Lyapunov exponent and the surface gravity \cite{Hashimoto:2016dfz}. Consequently, it is insightful to express our findings in terms of the surface gravity in BH spacetimes, particularly when examining specific limits. The surface gravity for this charged BH spacetime can be determined using the formula given,
\begin{equation}
    \kappa = \frac{-g^{'}_{tt}}{2\sqrt{-g_{tt}g_{rr}}} \vert_{r_{h}},
\end{equation}
which yields,
\begin{equation}
    \kappa_{RN_{AdS}} = \left( \frac{M}{r_{p}^{2}} - \frac{Q^{2}}{r_{p}^{3}} + \frac{r_{p}}{\ell^{2}}\right).
\end{equation}
with $M\geq Q$. Due to the complexity of the analytical calculations, we are unable to express the metric components in terms of the surface gravity. Consequently, the Lyapunov exponents and precession frequency are also not presented in terms of the surface gravity. In Fig. \ref{kappa}, we plot the surface gravity as a function of the charge parameter for different values of the cosmological constant to investigate the impact of the AdS geometry. It is clearly observed that the cosmological constant has a significant effect on the surface gravity, with the surface gravity decreasing as the cosmological constant increases. Notably, as the BH approaches extremality, the surface gravity decreases rapidly for all values of the cosmological constant. This indicates that the AdS geometry indeed has a notable influence on the QNMs in the presence of the charge parameter.
For massive BHs, the radius of photon sphere becomes,
\begin{equation}
    r_{pm+} = 3M \left( 1-\frac{2Q^2}{9M^2}\right).
\end{equation}
Therefore, the Penrose limit metric components are given as follow,
\begin{align}
    A_{11} &= \frac{4}{r_{pm}^{2}} -\frac{8(r_{pm}^{4} +\ell^{2}r_{pm}^{2}-2M\ell^{2}r_{pm}+Q^{2}\ell^{2})[r_{pm}(12r_{pm}^{2}+2\ell^{2})-4r_{pm}^{3}+2\ell^{2}r_{pm}-2M\ell^{2}]}{r_{pm}^{3}(4r_{pm}^{3}+2\ell^{2}r_{pm}-2M\ell^{2})^{2}} \\
    A_{22} & = -\frac{2}{r_{pm}^{2}} + \frac{8(r_{pm}^{4} +\ell^{2}r_{pm}^{2}-2M\ell^{2}r_{pm}+Q^{2}\ell^{2}) [4r_{pm}^{3}+2\ell^{2}r_{pm}-2M\ell^{2}-2r_{pm}]}{r_{pm}^{3}(4r_{pm}^{3}+2\ell^{2}r_{pm}-2M\ell^{2})^{2}}.
\end{align}
The $Q$ dependent terms are responsible for the different values of $A_{ij}$ in this particular charged BH. Now, the Laypunov exponent and precision frequency can be obtained by using the given formula as,
\begin{equation}
    \lambda^{2} = \frac{\Delta(r_{0}) [12\Delta(r_{0} - r_{0}^2 \Delta^{''}(r_{0}) ]}{2 r_{0}^{6}} \And \omega_{orb} = \frac{\Delta(r_{0})}{r_{0}^{4}}.
\end{equation}
which yields,
\begin{equation}
     \lambda^{2} = \frac{(r_{p}^{4}+\ell^{2}r_{p}^{2}-2M\ell^{2}r_{p}+Q^{2}\ell^{2})(10\ell^{2}r_{p}^{2}-24M\ell^{2}r_{p}+12Q^{2}\ell^{2})}{2r_{p}^{6}},
\end{equation}
and
\begin{equation}
   \omega_{orb} =  \frac{\left( r_{p}^{4}+\ell^{2}r_{p}^{2}-2M\ell^{2}r_{p}+Q^{2}\ell^{2} \right)}{r_{p}^{4}}.
\end{equation}
The Lyapunov exponent and precession frequency are closely related to the stability and oscillatory nature of QNMs. The Lyapunov exponent helps to understand the damping and stability of the modes, while the precession frequency influences the resonant frequencies and the overall dynamical response of the BH system. In Fig. \ref{LvsQ}, we plot the Lyapunov exponents and the precession frequency for a constant value of the cosmological constant to examine their behavior as functions of the charge parameter. Our observations reveal that the Lyapunov exponent initially reaches a maximum at minimum value of the charge parameter, then decreases, and subsequently increases again as the BH approaches extremality. In contrast, the precession frequency exhibits a consistent trend, steadily increasing with the charge parameter.
\section{Quasinormal Modes} \label{sec6}
In order to compute QNMs, we can write the solution of massless scalar wave equation for the metrics components as,
\begin{equation} \label{psisol}
    \psi = e^{i(p_{v}v+p_{u}u \psi_{1}(x_{1}) \psi_{2}(x_{2}))}.
\end{equation}
We separate the solution and obtain two independent equations for $\psi_{1}(x_{1})$ and  $\psi_{2}(x_{2})$ as follow,
\begin{align}
    \frac{1}{2p_{v}^2} \psi_{1}^{11}(x_{1}) + \frac{1}{2} A_{11} x_{1}^{2} \psi_{1}(x_{1}) &= \left( \frac{p_{u}}{2p_{v}} +C_{0} \right) \psi_{1} (x_{1}), \\
     \frac{1}{2p_{v}^2} \psi_{2}^{11}(x_{2}) + \frac{1}{2} A_{22} x_{2}^{2} \psi_{2}(x_{2}) &= \left( \frac{p_{u}}{2p_{v}} +C_{0} \right) \psi_{2} (x_{2}).
\end{align}
Here, $C_{0}$ is the separation constant. These two independent equations are generally solved in terms of parabolic cylinder functions. In order to compute quasinormal modes, we require an outgoing boundary condition in the unstable direction $x_{1}$, and a decaying boundary condition in the stable direction $x_{2}$. Together, these conditions lead to the quantization condition, 
\begin{equation}
    p_{u} = i\sqrt{A_{11}} \left( \frac{1}{2} + n_{1} \right) - \sqrt{A_{22}} \left( \frac{1}{2} + n_{2} \right),
\end{equation}
and $p_{v}$ is fixed from periodicity of $\psi$ in Eq. \eqref{psisol},
\begin{equation}
    p_{v} = L\omega_{orb}.
\end{equation}
We can thus obtain the quasinormal modes from Penrose limit using the given formula \cite{Ferrari:1984zz},
\begin{equation} \label{qnmformula}
    \omega_{Ln_{1}n_{2}} = L \omega_{orb} + \left( \frac{1}{2} + n_{2} \right) \omega_{prec} - i\lambda \left( \frac{1}{2} + n_{1} \right).
\end{equation}
Therefore, for RN BH the formula in Eq. \eqref{qnmformula} yields,
\begin{equation} \label{FEQNM}
     \omega_{Ln_{1}n_{2}} = \frac{L\Delta(r_{p})}{r_{p}^{4}} + \left( \frac{1}{2} + n_{2} \right) \left( \frac{\Delta(r_{p})\mathcal{F}(r_{p})}{2r_{p}^{6}} \right) - i \left( \frac{\Delta(r_{p})\mathcal{F}(r_{p})}{2r_{p}^{6}} \right)  \left( \frac{1}{2} + n_{1} \right).
\end{equation}
with, $\Delta({r_{p})=r_{p}^{4}+\ell^{2}r_{p}^{2}-2M\ell^{2}r_{p}+Q^{2}\ell^{2}}$ and $\mathcal{F}(r_{p}) = 10\ell^{2}r_{p}^{2}-24M\ell^{2}r_{p}+12Q^{2}\ell^{2}$.
The obtained expression best express the structure of the geometrical optics quasinormal
modes as it will emerge in general complicated form, but here it can of course be simplified using $\omega_{prec}$=$\omega_{orb}$ as,
\begin{equation}
    \omega_{ln} = \left( l + \frac{1}{2} \right) \frac{\Delta(r_{p})}{r_{p}^{4}} - i  \left( \frac{\Delta(r_{p})\mathcal{F}(r_{p})}{2r_{p}^{6}} \right)  \left( \frac{1}{2} + n \right).
\end{equation}
 The exponential decay of QNMs is linked to the exponential divergence of null geodesics. In other words, the Lyapunov exponents of null geodesics and the QNMs frequencies can be directly inferred from the plane wave Hamiltonian evaluated at the photon sphere. Using Eq. \ref{FEQNM}, we can investigate the behavior of QNMs by varying different BH parameters, such as charge, mass, and the cosmological constant. By plotting the real and imaginary parts of the QNMs as functions of these parameters, we can analyze how these parameters influence the frequencies and damping rates of the oscillatory modes.
 \begin{figure}[H]
	\begin{center}
        \includegraphics[width=.8\textwidth, height=6cm]{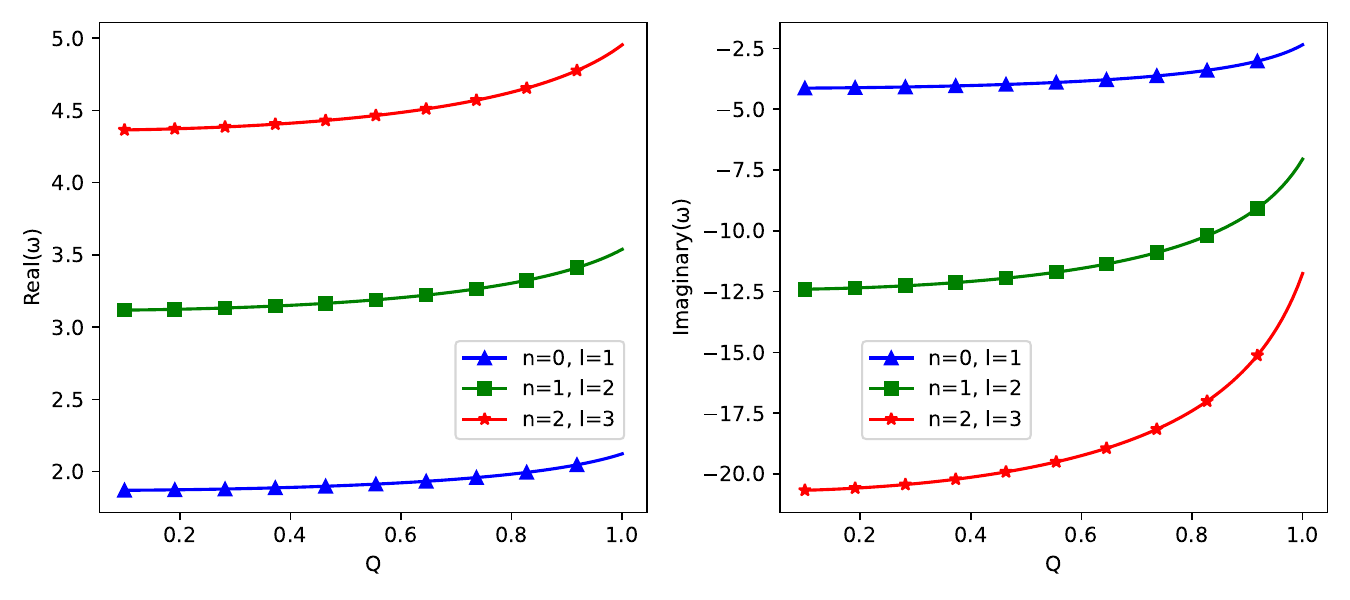} 
\end{center}
	\caption{The variation of real and imaginary parts of QNMs frequencies with charge parameter for different combination of $n$ and $l$ with $M=1$ and $\Lambda=-0.45$. } \label{qnm1}
\end{figure}
 \begin{figure}[H]
	\begin{center}
        {\includegraphics[width=.8\textwidth, height=6cm]{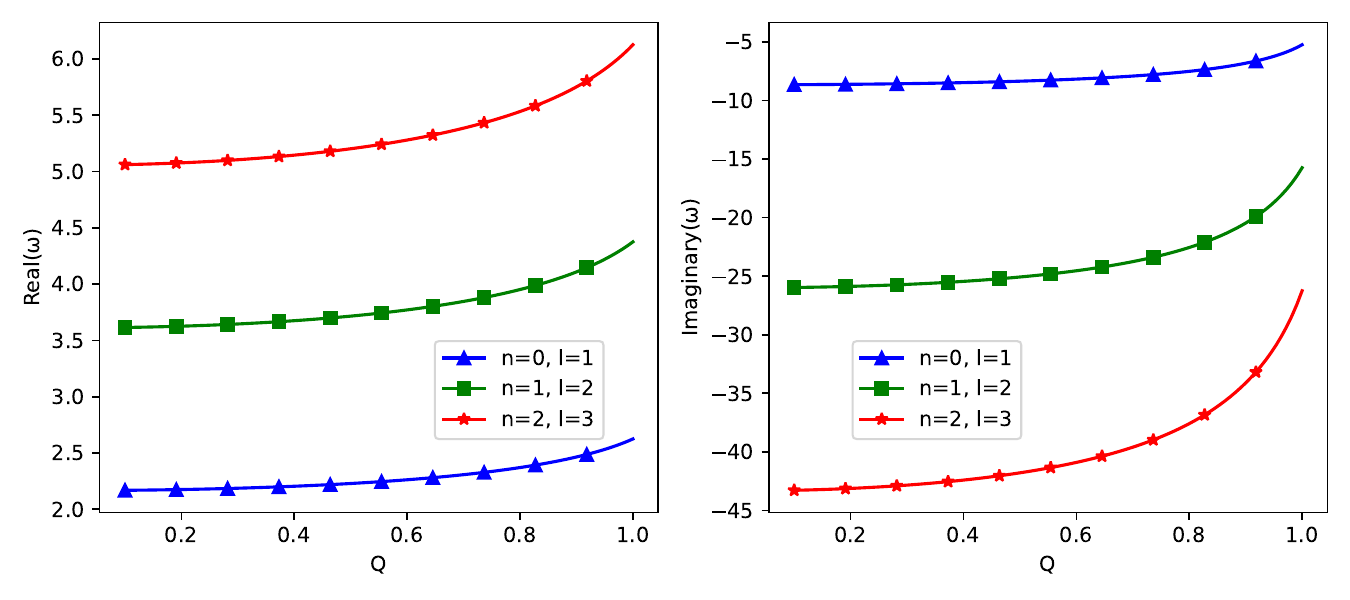}} 
\end{center}
	\caption{The variation of real and imaginary parts of QNMs frequencies with charge parameter for different combination of $n$ and $l$ with $M=1$ and $\Lambda=-0.25$. } \label{qnm2}
\end{figure}
\begin{figure}[H]
	\begin{center}
        {\includegraphics[width=.8\textwidth, height=6cm]{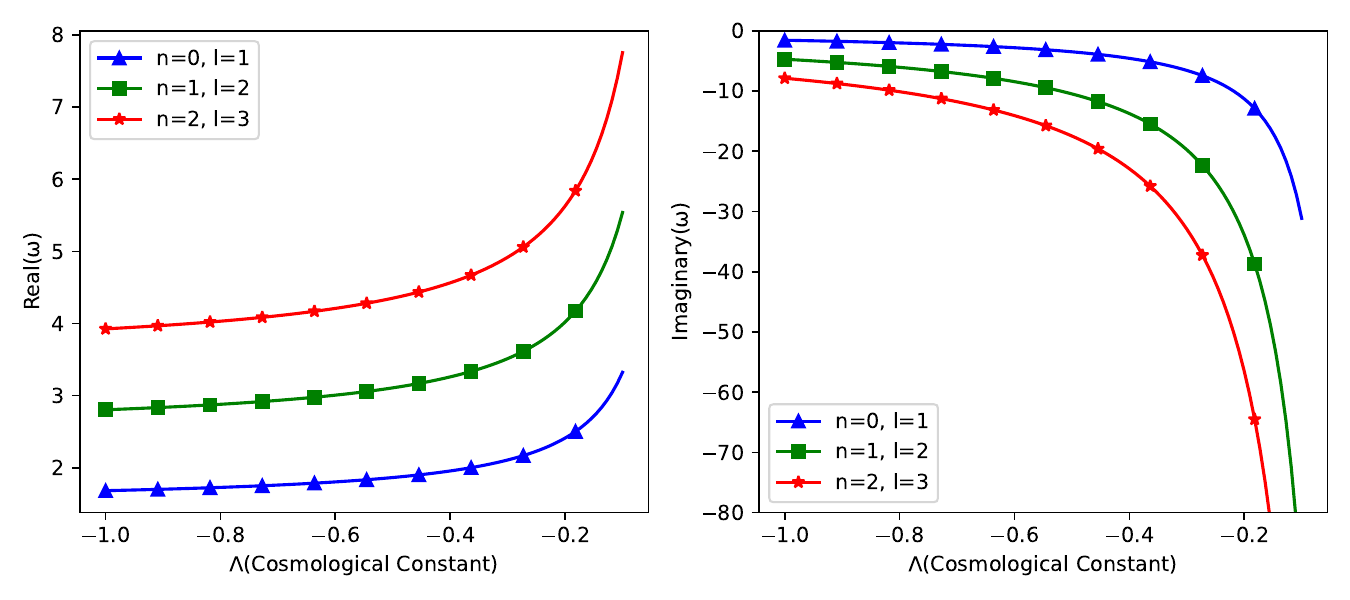}} 
\end{center}
	\caption{The variation of real and imaginary parts of QNMs frequencies with cosmological constant parameter for different combination of $n$ and $l$ with $M=1$ and $q=0.5$.} \label{qnm3}
\end{figure}
The graphical representation of the real and imaginary parts of the QNM frequencies is shown in Figures \ref{qnm1}-\ref{qnm3}. In Fig. \ref{qnm1}, we present these frequencies for different combinations of $l$ and $n$ as a function of the charge parameter $Q$, with a fixed cosmological constant $\Lambda=-0.45$. Similarly, Fig. \ref{qnm2} illustrates the same analysis for $\Lambda=-0.25$.
From these plots, we observe that both the real and imaginary parts of QNM frequencies increase as the charge parameter increases. However, the real frequency reaches its maximum for the highest values of $l$ and $n$, while the imaginary part reaches its maximum for the lowest values of $l$ and $n$. To further investigate the role of the cosmological constant on these frequencies, Fig. \ref{qnm3} presents the variation of real and imaginary QNM frequencies as a function of $\Lambda$ for a fixed charge parameter $Q$. We find that both frequencies tend to saturate at higher values of $\Lambda$. However, as $\Lambda$ decreases, the real part of the QNM frequencies increases sharply, whereas the imaginary part decreases, indicating that lower values of cosmological constants lead to longer-lived perturbations with higher oscillation frequencies. This behavior suggests that in asymptotically dS-like spacetimes, QNMs decay faster and exhibit lower oscillation frequencies, whereas in asymptotically AdS-like backgrounds, the system supports more oscillatory but longer-lived modes due to the stronger effective gravitational potential barrier. We have also presented the obtained results in tabular form for clarity and comparison. Table \ref{tab1} illustrates the QNM frequencies for a specific value of the cosmological constant, while Table \ref{tab2} represents the results in the absence of a cosmological constant. In both tables, we have used different values of the $Q$, $n$ and $l$ to systematically analyze their influence on the QNM frequencies. This tabular representation highlights the impact of the cosmological constant and charge parameter, facilitating a direct comparison of frequency shifts and decay rates.
A direct comparison between our results and those of Berti et al.\cite{berti2003quasinormal} and Wang et al.\cite{wang2000quasinormal}
is not feasible due to significant differences in methodology, parameter definitions and boundary conditions. These studies employ different computational techniques and normalization choices, leading to inherent variations in the QNM frequencies obtained. However, despite these differences, qualitative similarity is observed between our results and previous findings, providing an independent verification of our computations and reinforcing their reliability.
\begin{table}[H]
    \centering
    \renewcommand{\arraystretch}{1.3}
    \setlength{\tabcolsep}{8pt}
    \begin{minipage}{0.48\textwidth}
        \centering
        \begin{tabular}{||c||c|c|c|c||}
            \hline\hline
            $Q$ & $n$ & $l$ & Real($\omega_{ln}$) & Imag($\omega_{ln}$) \\
            \hline\hline
            \multirow{9}{*}{0.1}  
            & 0 & 1 & 1.870463  & -4.134502  \\ \cline{2-5}
            & 0 & 2 & 3.117439  & -4.134502  \\ \cline{2-5}
            & 0 & 3 & 4.364414  & -4.134502  \\ \cline{2-5}
            & 1 & 1 & 1.870463  & -12.403507 \\ \cline{2-5}
            & 1 & 2 & 3.117439  & -12.403507 \\ \cline{2-5}
            & 1 & 3 & 4.364414  & -12.403507 \\ \cline{2-5}
            & 2 & 1 & 1.870463  & -20.672511 \\ \cline{2-5}
            & 2 & 2 & 3.117439  & -20.672511 \\ \cline{2-5}
            & 2 & 3 & 4.364414  & -20.672511 \\ 
            \hline\hline
            \multirow{9}{*}{0.5}  
            & 0 & 1 & 1.903933  & -3.953242  \\ \cline{2-5}
            & 0 & 2 & 3.173221  & -3.953242  \\ \cline{2-5}
            & 0 & 3 & 4.442509  & -3.953242  \\ \cline{2-5}
            & 1 & 1 & 1.903933  & -11.859726 \\ \cline{2-5}
            & 1 & 2 & 3.173221  & -11.859726 \\ \cline{2-5}
            & 1 & 3 & 4.442509  & -11.859726 \\ \cline{2-5}
            & 2 & 1 & 1.903933  & -19.766210 \\ \cline{2-5}
            & 2 & 2 & 3.173221  & -19.766210 \\ \cline{2-5}
            & 2 & 3 & 4.442509  & -19.766210 \\ 
            \hline\hline
            \multirow{9}{*}{1.0}  
            & 0 & 1 & 2.123070  & -2.351684  \\ \cline{2-5}
            & 0 & 2 & 3.538451  & -2.351684  \\ \cline{2-5}
            & 0 & 3 & 4.953831  & -2.351684  \\ \cline{2-5}
            & 1 & 1 & 2.123070  & -7.055052  \\ \cline{2-5}
            & 1 & 2 & 3.538451  & -7.055052  \\ \cline{2-5}
            & 1 & 3 & 4.953831  & -7.055052  \\ \cline{2-5}
            & 2 & 1 & 2.123070  & -11.758420 \\ \cline{2-5}
            & 2 & 2 & 3.538451  & -11.758420 \\ \cline{2-5}
            & 2 & 3 & 4.953831  & -11.758420 \\ 
            \hline\hline
        \end{tabular}
        \caption{Computed values of $\omega_{ln}$ for different values of $Q$, $n$, and $l$, with $M = 1$ and $\Lambda = -0.45$.}
        \label{tab1}
    \end{minipage}
    \hfill
    \begin{minipage}{0.48\textwidth}
        \centering
        \begin{tabular}{||c||c|c|c|c||}
            \hline\hline
            $Q$ & $n$ & $l$ & Real($\omega_{ln}$) & Imag($\omega_{ln}$) \\
            \hline\hline
            \multirow{9}{*}{0.1}  
            & 0 & 1 & 1.905000  & -4.250000  \\ \cline{2-5}
            & 0 & 2 & 3.160000  & -4.250000  \\ \cline{2-5}
            & 0 & 3 & 4.415000  & -4.250000  \\ \cline{2-5}
            & 1 & 1 & 1.905000  & -12.750000 \\ \cline{2-5}
            & 1 & 2 & 3.160000  & -12.750000 \\ \cline{2-5}
            & 1 & 3 & 4.415000  & -12.750000 \\ \cline{2-5}
            & 2 & 1 & 1.905000  & -21.250000 \\ \cline{2-5}
            & 2 & 2 & 3.160000  & -21.250000 \\ \cline{2-5}
            & 2 & 3 & 4.415000  & -21.250000 \\ 
            \hline\hline
            \multirow{9}{*}{0.5}  
            & 0 & 1 & 1.930000  & -4.000000  \\ \cline{2-5}
            & 0 & 2 & 3.200000  & -4.000000  \\ \cline{2-5}
            & 0 & 3 & 4.470000  & -4.000000  \\ \cline{2-5}
            & 1 & 1 & 1.930000  & -12.000000 \\ \cline{2-5}
            & 1 & 2 & 3.200000  & -12.000000 \\ \cline{2-5}
            & 1 & 3 & 4.470000  & -12.000000 \\ \cline{2-5}
            & 2 & 1 & 1.930000  & -20.000000 \\ \cline{2-5}
            & 2 & 2 & 3.200000  & -20.000000 \\ \cline{2-5}
            & 2 & 3 & 4.470000  & -20.000000 \\ 
            \hline\hline
            \multirow{9}{*}{1.0}  
            & 0 & 1 & 2.200000  & -2.500000  \\ \cline{2-5}
            & 0 & 2 & 3.600000  & -2.500000  \\ \cline{2-5}
            & 0 & 3 & 5.000000  & -2.500000  \\ \cline{2-5}
            & 1 & 1 & 2.200000  & -7.500000  \\ \cline{2-5}
            & 1 & 2 & 3.600000  & -7.500000  \\ \cline{2-5}
            & 1 & 3 & 5.000000  & -7.500000  \\ \cline{2-5}
            & 2 & 1 & 2.200000  & -12.500000 \\ \cline{2-5}
            & 2 & 2 & 3.600000  & -12.500000 \\ \cline{2-5}
            & 2 & 3 & 5.000000  & -12.500000 \\ 
            \hline\hline
        \end{tabular}
        \caption{Computed values of $\omega_{ln}$ for different values of $Q$, $n$, and $l$, with $M = 1$ and $\Lambda = 0$.}
        \label{tab2}
    \end{minipage}
\end{table}
\section{Conclusions} \label{sec7}
\noindent We investigated the QNMs of RN AdS BH using the Penrose process method. Our results show that the real and imaginary parts of the QNM frequencies increase with the charge parameter \(Q\), with the real part maximizing for higher values of angular momentum and higher overtones, while the imaginary part is dominant for lower \(l, n\). Additionally, as the cosmological constant \(\Lambda\) decreases, the real frequency increases sharply, while the imaginary frequency decreases, indicating longer-lived perturbations. Furthermore, the QNM frequencies saturate at higher values of \(\Lambda\), suggesting a stabilizing effect of the cosmological constant on the BH perturbations. The results demonstrate that asymptotically de Sitter (dS) spacetimes lead to faster-decaying, low-frequency QNMs, while asymptotically anti-de Sitter (AdS) spacetimes support longer-lived, higher-frequency oscillations. We compared our results with those of Wang et al.\cite{wang2000quasinormal} and Berti et al.\cite{berti2003quasinormal}, which are based on the method proposed by Horowitz et al.~\cite{Horowitz:1999jd} for AdS spacetimes. Although there are differences due to variations in methodology, parameter definitions, and boundary conditions, we found qualitative agreement with their findings. This consistency reinforces the validity of our approach and confirms the applicability of the Penrose process method for studying charged BHs in AdS geometry.
\section{Acknowledgments}
The authors SK and HN acknowledge the use of facilities at ICARD, Gurukula Kangri (Deemed to be University), Haridwar, India. The author, SK, sincerely acknowledges IMSc for providing exceptional research facilities and a conducive environment that facilitated his work as an Institute Postdoctoral Fellow. One of the authors, HN., would also like to thank IUCAA, Pune, for the support under its associateship program, where a part of this work was done. The author HN also acknowledges the financial support provided by the Science and Engineering Research Board (SERB), New Delhi, through grant number CRG/2023/008980. 
 AR would like to thank Kwinten Fransen and Vishnu Jejjala for the valuable discussions. The work of AR has been
supported by the NRF-TWAS doctoral Grant MND190415430760, UID 121811 and the DSI/NRF South African Research Chairs Initiative (SARChI) Research Chair in Theoretical Particle Cosmology, grant number 78554. 

\bibliography{references.bib}

\end{document}